\documentclass[aps,prl,showpacs,twocolumn,floats]{revtex4}
\usepackage{amsfonts,amssymb,amsmath}
\usepackage{graphics}
\usepackage{longtable}
\usepackage{subfigure}
\usepackage{epsfig}
\begin{document}
\title{The Varieties of Dynamic Multiscaling in Fluid Turbulence}   
\author{Dhrubaditya Mitra} 
\author{ Rahul Pandit}
\altaffiliation[Also at ]{Jawaharlal Nehru Centre For Advanced
Scientific Research, Jakkur, Bangalore, India}
\affiliation{Centre for Condensed Matter Theory, 
Department of Physics, Indian Institute of Science, 
Bangalore 560012, India}
\begin{abstract}
We show that different ways of extracting time scales from time-dependent
velocity structure functions lead to different dynamic-multiscaling 
exponents in fluid turbulence. These exponents are related to equal-time 
multiscaling exponents by different classes of bridge relations which we
derive. We check this explicitly by detailed numerical simulations of 
the GOY shell model for fluid turbulence. Our results can be generalized to 
any system in which both equal-time and time-dependent structure 
functions show multiscaling.  
\end{abstract}
\keywords{Turbulence, Multifractality, Dynamic Scaling}
\pacs{47.27.i, 47.53.+n }
\maketitle

The dynamic scaling of time-dependent correlation functions in the 
vicinity of a critical point was understood soon after the 
scaling of equal-time correlations~\cite{hoh77}. By contrast, the 
development of an understanding of the dynamic mutiscaling of 
time-dependent velocity structure functions in homogeneous, isotropic
fluid turbulence is still continuing; and studies of it lag far 
behind their analogs for the multiscaling of equal-time velocity
structure functions~\cite{fri96}. There are three major reasons for this:
(1) The multiscaling of equal-time velocity structure functions in fluid 
turbulence is far more complex than the scaling of equal-time correlation 
functions in critical phenomena~\cite{fri96}.
(2) The dynamic scaling of Eulerian-velocity structure functions is
dominated by sweeping effects that relate temporal and spatial scales 
linearly and thus lead to a trivial dynamic-scaling exponent
$z_{\cal{E}} = 1$, where the subscript $\cal{E}$ stands for Eulerian.
(3) Even if this dominant temporal scaling because of sweeping effects is 
removed (see below), time-dependent velocity structure
functions do not have simple scaling forms. As has been recognized in 
Ref.~\cite{lvo97}, in the fluid-turbulence context, 
an infinity of dynamic-multiscaling exponents is required. These are 
related to the equal-time multiscaling exponents by {\it bridge 
relations}, one class of which were obtained in Ref.~\cite{lvo97}. In
the forced-Burgers-turbulence context 
a few bridge relations of another class were obtained in Refs.~\cite{hay98}
and ~\cite{hay00}. 
If the bridge relations of 
Refs.~\cite{lvo97} and \cite{hay98,hay00} are compared naively, then they 
disagree with each other. 
However, the crucial point about dynamic multiscaling, 
not enunciated clearly hitherto, though partially implicit in 
Refs.~\cite{lvo97,hay98,hay00}, is that {\it different ways of extracting time 
scales} from time-dependent velocity structure functions yield 
{\it different dynamic-multiscaling exponents} that are related to the 
equal-time multiscaling exponents by {\it different classes of bridge 
relations}. We systematize such bridge relations by distinguishing three types 
of methods that can be used to extract time scales; these are based, 
respectively, on {\it integral} $I$, {\it derivative} $D$, and 
{\it exit-time} $E$ scales. We then derive the bridge relations 
for dynamic-multiscaling exponents for these three methods. Finally we check 
by an extensive numerical simulation that such bridge relations are  
satisfied in the GOY shell model for fluid turbulence. 

To proceed further let us recall that in homogeneous, isotropic 
turbulence, the equal-time, order-$p$, velocity structure function 
${\mathcal S}_p(\ell) \equiv \langle [\delta v_{\parallel}({\vec x},t,\ell)]^p \rangle
\sim \ell^{\zeta_p} $, for $\eta_d \ll \ell \ll L$, where 
$\delta v_{\parallel}({\vec x},t,\ell) = [{\vec v}({\vec x} + {\vec \ell} ,t) 
                                - {\vec v}({\vec x} ,t ) ] \cdot  
                                ({\vec \ell}/\ell)$, 
${\vec v}({\vec x},t)$ is the fluid velocity at point ${\vec x}$ and
time $t$, $L$ is the large spatial scale at which energy is injected into
the system, $\eta_d$ is the small length scale at which viscous 
dissipation becomes significant, $\zeta_p$ is the order-$p$, equal-time 
multiscaling exponent, and the angular brackets denote an average over 
the statistical steady state of the turbulent fluid.
The 1941 theory (K41) of  Kolmogorov \cite{kol41} yields the simple 
scaling result $\zeta_p^{K41} = p/3$. However, experiments and simulations
indicate multiscaling, i.e., $\zeta_p$ is a nonlinear, convex function 
of $p$; and the von-K\'arm\'an-Howarth relation \cite{fri96} yields
$\zeta_3 = 1$.  To study dynamic multiscaling we use the 
longitudinal, time-dependent, order-$p$ structure function~\cite{lvo97} 
\begin{eqnarray}
{\mathcal F}_p(\ell,\{t_1,\ldots,t_p\}) \equiv 
        \langle [\delta v_{\parallel}({\vec x},t_1,\ell) \ldots  
                 \delta v_{\parallel}({\vec x},t_p,\ell)] \rangle .
\label{dynsp}
\end{eqnarray}
Clearly ${\mathcal F}_p(\ell,\{t_1= \ldots =t_p=0\})={\mathcal S}_p(\ell)$. 
We normally restrict ourselves to the simple case $t_1=t_2=\ldots = t_q 
\equiv t$ 
and $t_{q+1}=t_{q+2} = \ldots = t_p = 0$, for notational simplicity
write ${\mathcal F}_p(\ell,t)$, and suppress the $q$ dependence which
should not affect dynamic-multiscaling exponents (see below). To remove 
the sweeping effects mentioned above, we must of course use 
quasi-Lagrangian~\cite{lvo97,bel88} or Lagrangian~\cite{kan99} 
velocities in Eq. (\ref{dynsp}), but we do not show this explicitly here 
for notational convenience. Given ${\mathcal F}_p(\ell,t)$, we can 
extract a characteristic time scale $\tau_p(\ell)$ in several different 
ways, as we describe later. The dynamic-multiscaling ansatz 
$ \tau_p(\ell) \sim \ell^{z_p}$ can now be used to determine the order-$p$ 
{\it dynamic-multiscaling exponents} $z_p$.  
Furthermore, a naive extension of K41 to dynamic 
scaling~\cite{mit03} yields $z_p^{K41} = \zeta_2^{K41} = 2/3$ for all $p$.

In the multifractal model \cite{fri96} the velocity of a turbulent flow is 
assumed to possess a range of universal scaling exponents 
$h \in {\cal I} \equiv (h_{min},h_{max})$. For each $h$ in this range, 
there exists a set $\Sigma_h \subset \mathbb{R}^3$ of fractal dimension $D(h)$, 
such that 
$ \frac{\delta v({\vec{r}},\ell)}{v_L} \propto (\frac{\ell}{L})^h $
  for $ \vec{r} \in {\Sigma}_h$ ,
with $v_L$ the velocity at the forcing scale $L$, whence
\begin{eqnarray}
 \frac{{\mathcal S}_p(\ell)}{v_L^p} \equiv \frac{\langle \delta v^p(\ell) 
   \rangle} {v_L^p}
   \propto \int_{{\cal I}} d\mu(h)(\frac{\ell}{L})^{{\cal Z}(h)} \/,
\label{eq-sp}
\end{eqnarray} 
where ${\cal Z}(h) = [ph + 3 - D(h)]$, the measure $d\mu(h)$ gives the 
weight of the fractal sets, and a saddle-point evaluation of the 
integral yields $\zeta_p = \underset{h}{\inf}  [{\cal Z}(h)]$.
The $ph$ term in ${\cal Z}(h)$ comes from $p$ factors of $(\ell/L)$ in 
Eq. (\ref{eq-sp}); the $3-D(h)$ term comes from an additional factor of
$(\ell/L)^{3-D(h)}$, which is the probability of being within a 
distance $\sim \ell$ of the set $\Sigma_h$ of dimension $D(h)$
that is embedded in three dimensions. 
 Similarly for the time-dependent structure function  
\begin{eqnarray}
 \frac{{\mathcal F}_p(\ell,t)}{v_L^p} \propto \int_{{\cal I}}d\mu(h)
 (\frac{\ell}{L})^{{\cal Z}(h)}{\cal G}^{p,h}(\frac{t}{\tau_{p,h}}),
\label{mcalf}
\end{eqnarray} 
where ${\cal G}^{p,h}(\frac{t}{\tau_{p,h}})$ has a characteristic decay 
time $\tau_{p,h} \sim \ell/\delta v(\ell) \sim \ell^{1 - h}$, and 
${\cal G}^{p,h}(0) = 1$. 
If $\int_0^{\infty} t^{(M-1)} {\cal G}^{p,h} dt $ exists, 
we can define the order-$p$, degree-$M$, {\it integral} time scale
\begin{eqnarray}
 {\cal T}^I_{p,M}(\ell) \equiv 
 \biggl[ \frac{1}{{\mathcal S}_p(\ell)}
\int_0^{\infty}{\mathcal F}_p(\ell,t)t^{(M-1)} dt
\biggl]^{(1/M)} .
\label{timp} 
\end{eqnarray} 
We can now define the {\it integral} dynamic-multiscaling exponents $z^I_{p,M}$
 via 
${\cal T}^I_{p,M} \sim \ell^{z^I_{p,M}}$.
By substituting the multifractal form (\ref{mcalf}) in 
Eq. (\ref{timp}),  computing the time integral 
first, and then performing the integration over
the multifractal measure by the saddle-point method, we obtain the 
{\it integral} bridge relations 
\begin{eqnarray}
z^I_{p,M} = 1 + [\zeta_{p-M} - \zeta_p]/M ,
\label{zipm}
\end{eqnarray}
which was first obtained in Ref.~\cite{lvo97}.
Likewise, if $\frac{\partial^M}{\partial t^M}{\cal G}^{p,h} |_{t=0} $ 
exists, we can define the order-$p$, degree-$M$, {\it derivative} time scale 
\begin{eqnarray}
 {\cal T}^D_{p,M} \equiv \biggl[\frac{1}{{\mathcal S}_p(\ell)}
                   \frac{\partial^M}{\partial t^M} 
                  {\mathcal F}_p(\ell,t) \biggl|_{t=0} \biggl]^{(-1/M)} , 
\label{tdpm}
\end{eqnarray} 
the {\it derivative} dynamic-multiscaling exponents $z^D_{p,M}$ via 
${\cal T}^D_{p,M} \sim \ell^{z^D_{p,M}}$, and thence obtain the 
{\it derivative} bridge relation
\begin{eqnarray}
z^D_{p,M} = 1 + [\zeta_p - \zeta_{p+M}]/M .
\label{zdpm}
\end{eqnarray}
Such derivative bridge relations, for the special cases (a) $p = 2, M = 1$ and 
(b) $p = 2, M = 2$, were first obtained in the forced-Burgers-turbulence
context in Refs.~\cite{hay00} and~\cite{hay98}, respectively~\cite{foot-jh}, 
without using quasi-Lagrangian velocities but by
using other methods to suppress sweeping effects. Case (a) yields the
interesting result $z^D_{2,1} = \zeta_2$, since $\zeta_3 = 1$.
Both relations (\ref{zipm}) and (\ref{zdpm}) reduce to 
$z_p^{K41} = 2/3$ if we assume K41 scaling for the equal-time structure 
functions.  

If we consider $n$ non-zero time arguments for the structure function, 
${\mathcal F}_{p,n}(\ell,t_1,\ldots ,t_n,\ldots, 0 \ldots, 0)$, 
which we denote by ${\mathcal F}_{p,n}(\ell,t_1,\ldots ,t_n )$ for
notational simplicity, we can define the 
integral time scale,   
$ {\cal T}^I_{p,M,n}(\ell) \equiv [\frac{1}{{\mathcal S}_p(\ell)} \int_0^{\infty} 
        {\mathcal F}_p(\ell,t_1,\ldots ,t_n) 
         t_1^{m_1-1}dt_1 \ldots t_n^{m_n-1}dt_n ]^{1/(Mn)},$
and the derivative time scale,  
$ {\cal T}^D_{p,M,n}(\ell) \equiv 
              [\frac{1}{{\mathcal S}_p(\ell)}
			  \frac{\partial^{m_1}}{\partial t_1^{m_1}} \cdots  
                           \frac{\partial^{m_n}}{\partial t_n^{m_n}} 
                            {\mathcal F}_p(\ell,t_1, \ldots ,t_n) 
                            |_{t_1=0,\ldots ,t_n=0}]^{-1/(Mn)} $,
where $M = \sum_{i=1}^n m_i $.   
From these we can obtain, as above, two generalized bridge relations : 
\begin{eqnarray}
z^I_{p,M,n} = 1 + (\zeta_{p-nM} - \zeta_p)/(nM) ; \nonumber \\
z^D_{p,M,n} = 1 + (\zeta_p - \zeta_{p+nM})/(nM) .
\label{zpgen}
\end{eqnarray}

In the rest of this paper, we study time-dependent structure functions of
the GOY shell model for fluid turbulence 
\cite{fri96,goy73,kad95}: 
\begin{eqnarray}
 (\frac{d}{dt} + \nu k_n^2)u_n = i(a_n u_{n+1}u_{n+2} + \nonumber \\ 
 \hspace{1cm}  b_n u_{n-1}u_{n+1} + c_n u_{n-1}u_{n-2} )^{\ast} + f_n .
\label{goy}
\end{eqnarray}
Here the complex, scalar velocity $u_n\/$, for the shell $n\/$, depends on 
the one-dimensional, logarithmically spaced wavevectors $k_n = k_0 2^n\/$, 
complex conjugation is denoted by $\ast$, and the coefficients $a_n = k_n$,
$b_n = -\delta k_{n-1}$, and $c_n = -(1-\delta)k_{n-2}$, with $\delta=1/2$,
are chosen to conserve the shell-model analogs of energy and helicity 
in the inviscid, unforced limit. By construction, the velocity in a 
given shell is affected directly only by velocities in nearest- and
next-nearest-neighbor shells. By contrast, all Fourier modes of the 
velocity field interact with each other in the Navier-Stokes equation as can 
be seen easily by writing it in wave-vector space. Thus the GOY shell model 
does not have the sweeping effect by which modes (eddies) corresponding to 
the largest length scales affect all those at smaller length scales 
{\it directly}. Hence it has been suggested that the GOY shell model should 
be thought of as a model for quasi-Lagrangian velocities~\cite{bif99}. We might 
anticipate therefore that GOY-model structure functions should
not have the trivial dynamic scaling associated with Eulerian velocities; we 
show this explicitly below.  
We integrate the GOY model (\ref{goy})
by using the slaved, Adams-Bashforth scheme ~\cite{dha97,pis93}, and $22$ 
shells ($1 \leq n \leq 22$), with $f_n = 0\/$ for $n \geq 2\/$ and 
$f_1 = (1+i)\times5\times10^{-3}$(Table~\ref{para-table}).
The equal-time structure function of order-$p$  and the associated
exponent is defined by 
 $S_p(k_n) = \langle |u_n|^p \rangle \sim k_n^{-\zeta_p}$.
However, the static solution of Eq. (\ref{goy}) exhibits a 3-cycle with the
shell index $n$, which is effectively filtered out~\cite{kad95} if we use 
$\Sigma_p(k_n) \equiv \langle |\Im(u_{n+2}u_{n+1}u_n - 
                  (1/4)u_{n-1}u_nu_{n+1})|^{p/3} \rangle \sim k_n^{-\zeta_p},$ 
 to determine
$\zeta_p$. These exponents are in close agreement with those found for  
homogeneous, isotropic fluid turbulence in three dimension~\cite{kad95}.   
Data for the exponents $\zeta_p$ from our calculations are given in 
Table~\ref{data-table}. We analyse the velocity $(u_n(t))$ 
time-series for $ n = 4$ to $13$, which corresponds to wave-vectors well 
within the inertial range.
The smaller the wave-vector $k_n$ the slower is the evolution of $u_n(t)$, so
it is important to use different temporal sampling rates for velocities
in different shells. We use sampling rates of $50\times \delta t$ for 
$ 4 \le n \le 8$ and  $10 \times \delta t$ for  $9 \le n \le 13$,
respectively.  
\begin{table}
\framebox{\begin{tabular}{c|c|c|c|c|c|c|c|c}
$\nu$ & $\delta t$ & $\lambda$ & $u_{rms}$ &  $Re_{\lambda}$ & $L_{int}$ &
$\tau_L$ & $T_{tr}$ & $T_{av}$  \\
\hline
$10^{-7}$ & $2 \times 10^{-4}$ & $0.7 $ & $0.35 $ & $ 2 \times 10^{6}$  & 
$6.3$ & $10^5 \delta t$ & $ 5 \times 10^4 \delta t$ &  $10^5 \tau_L$   \\
\end{tabular}}
\caption{Viscosity $\nu$, the time-step $\delta t$, Taylor microscale 
$\lambda \equiv (\sum_n |u_n|^2/k_n /\sum_n k_n |u_n|^2)^{1/2}$, 
the root-mean-square velocity  
$u_{rms} \equiv [2\sum_n |u_n|^2/k_n/(2\pi k_1)]^{1/2}$, 
the Taylor-microscale Reynolds number 
$Re_{\lambda}\equiv \lambda u_{rms}/\nu$, 
the integral scale  
$L_{int} \equiv (\sum_n |u_n|^2/k_n^2)/(\sum_n |u_n|^2/k_n ) $,  
and the box-size eddy turnover time 
$\tau_L \equiv L_{int}/u_{rms}$, 
that we use in our numerical simulation of the GOY shell model.
Data from the first $T_{tr}$ time steps are discarded so that transients 
can die down. We then average our data for time-dependent structure functions 
for an averaging time $ T_{av}$. 
}
\label{para-table}
\end{table}

\begin{figure*}
\includegraphics[height=3.2cm]{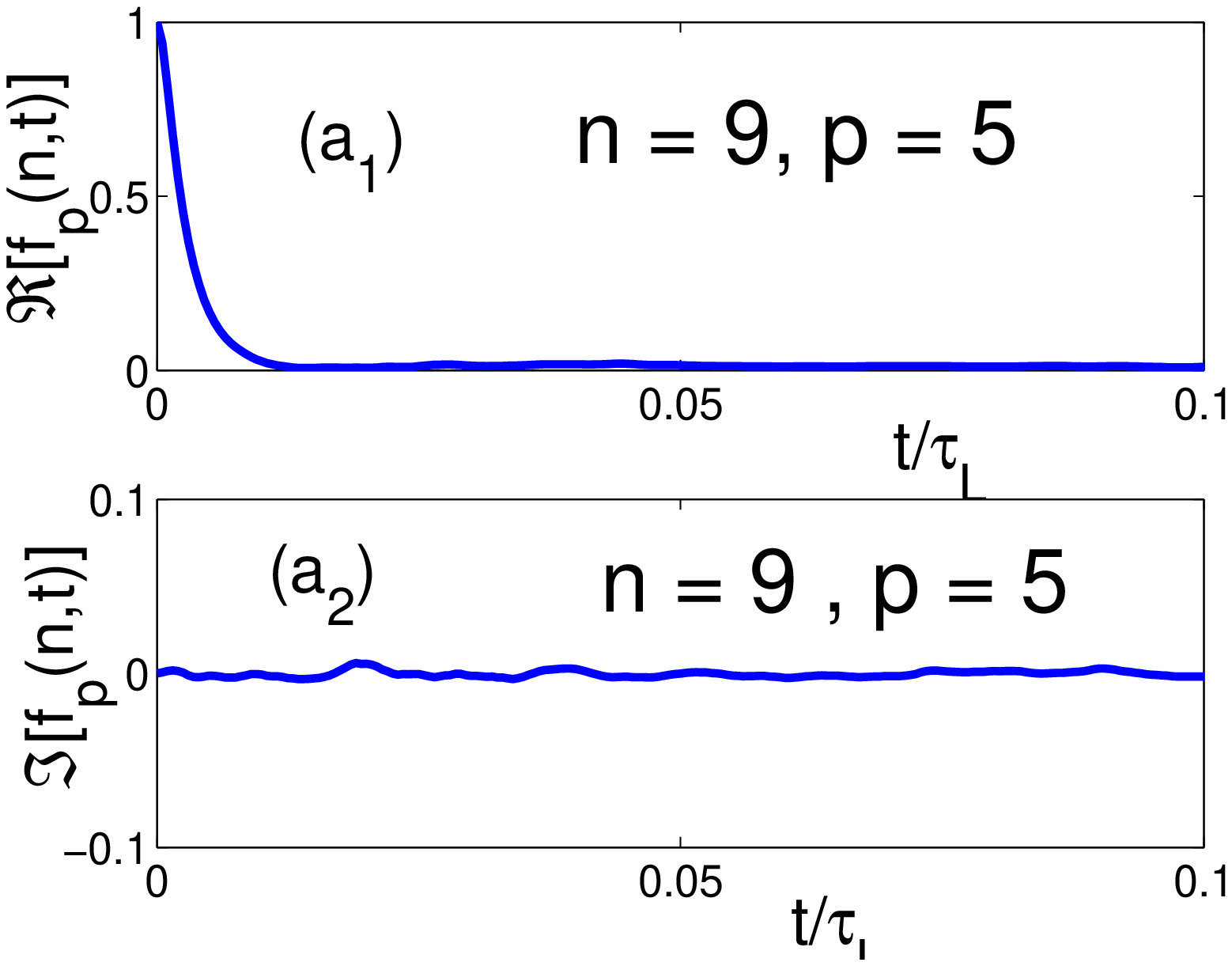}
\includegraphics[height=3.2cm]{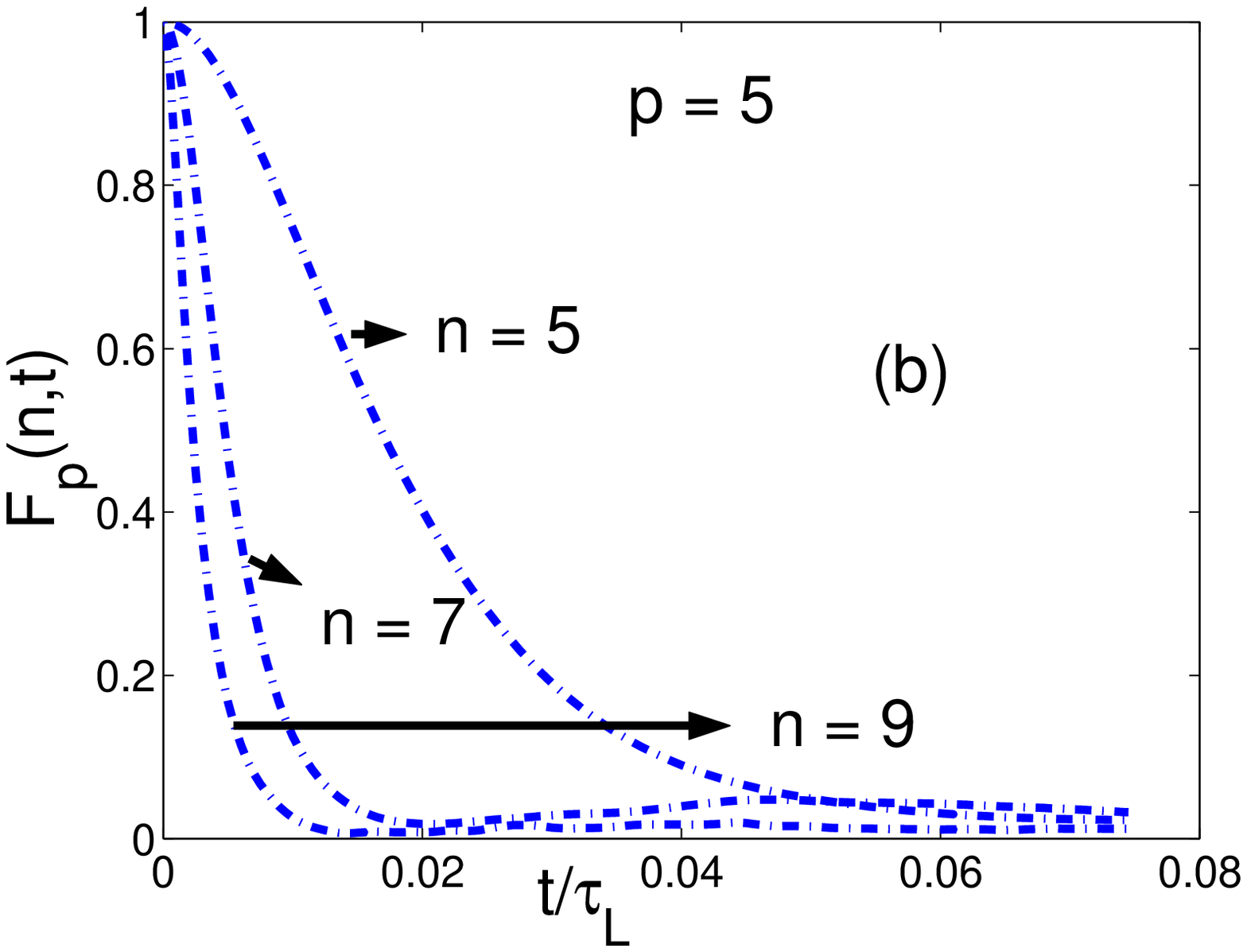}  
\includegraphics[height=3.2cm]{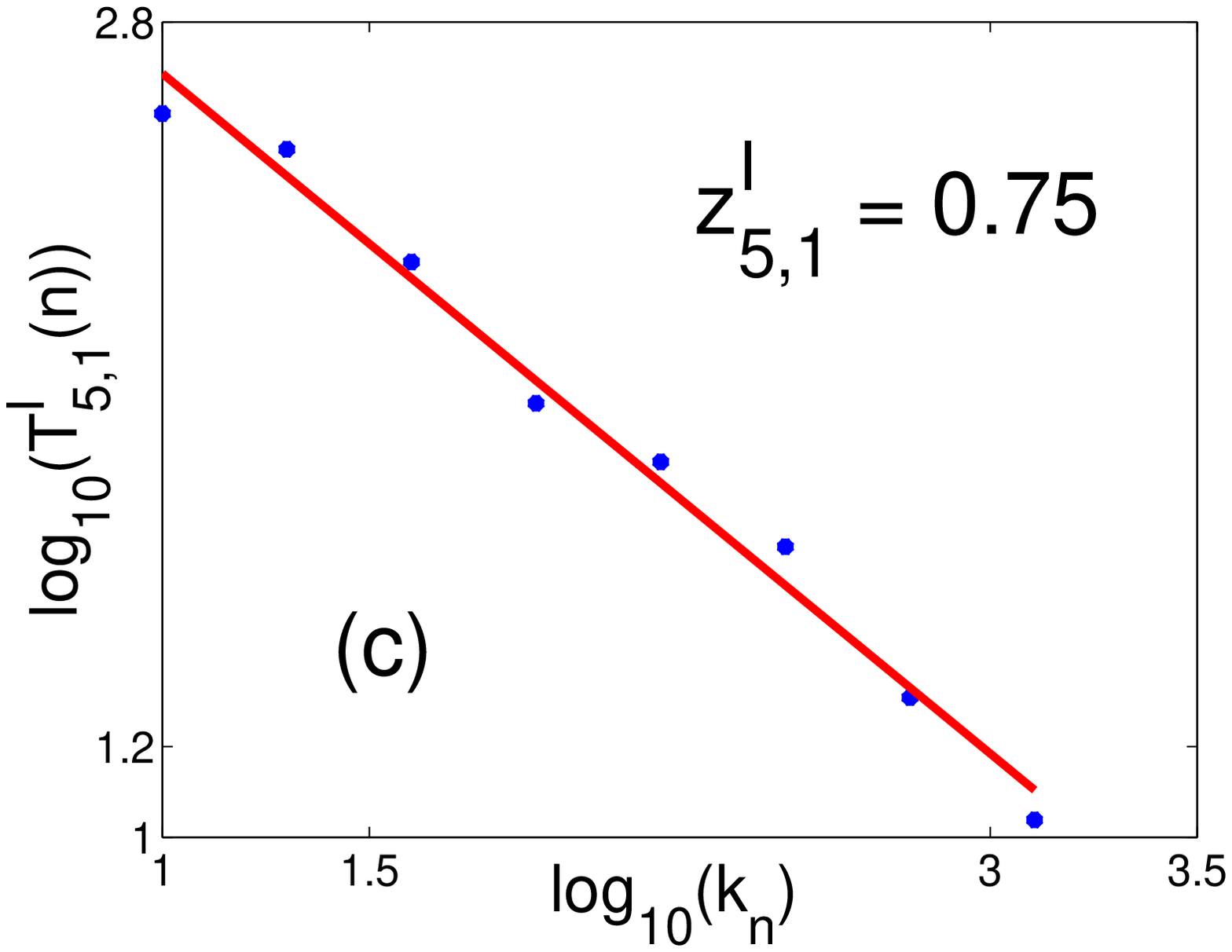}
\includegraphics[height=3.2cm]{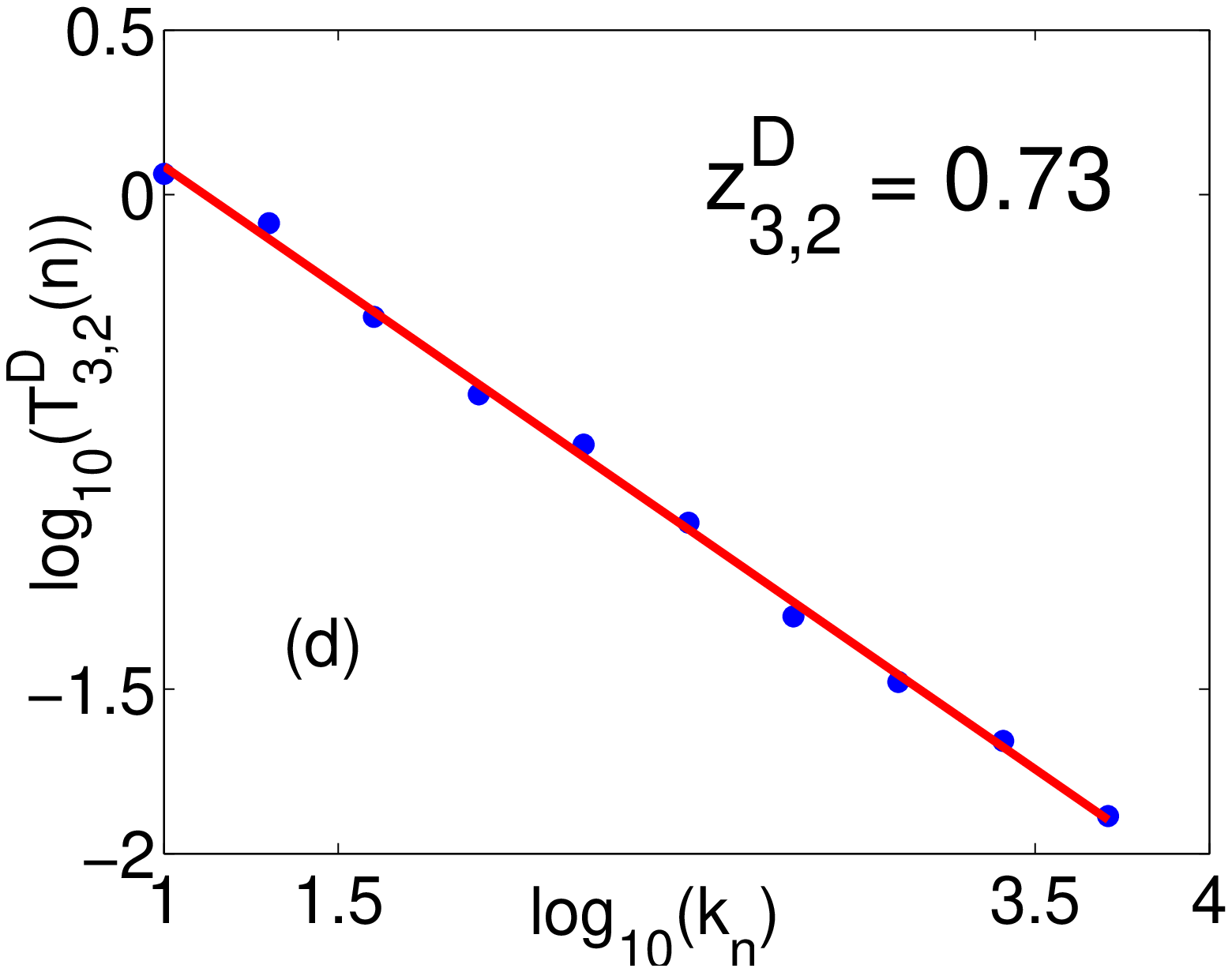} \\ 
\caption{\small Plots of real (${\text a_1}$) and imaginary (${\text a_2}$)
parts of the time-dependent structure function $f_p(n,t)$ for the GOY shell 
model for shell number $n=9$ and order $p=5$ versus time $t/\tau_L$, 
where $\tau_L$ is the box-size eddy turnover time (Table~\ref{para-table}).
Note that $\Im[f_p(n,t)]$ is negligibly small compared to 
$F_p(n,t) = \Re[f_p(n,t)]$. (b) $F_p(n,t)$ versus $t/\tau_L$ for 
$p=5$ and $n=5,7,$ and $9$. Representative log-log plots (base 10)
of the integral (c) and derivative (d) time scales $T^I_{5,1}(n)$
and $T^D_{3,2}(n)$ versus $k_n$; the slopes of the linear least-square fits
in (c) and (d) yield the dynamic exponents $z^I_{5,1}$ and $z^D_{3,2}$, 
respectively.     
}
\label{cofun}
\end{figure*}

For the GOY shell model we use the normalized, order-$p$, complex,  
time-dependent structure function,  
$  f_p(n,t) \equiv \langle [u_n(0)u_n^*(t)]^{p/2}\rangle /S_p(k_n)$ , 
which has both real and imaginary parts. The representative plot of
Fig.~\ref{cofun} shows that the imaginary part of $f_p(n,t)$ is 
negligibly small compared to its real part.  Hence we work with the 
real part of $ f_p(n,t)$, i.e.,  $ F_p(n,t) \equiv \Re [f_p(n,t)] $. 

Integral and derivative time scales can be defined for the 
shell model (\ref{goy}) as in Eqs. (\ref{timp}) and (\ref{tdpm}). 
We now concentrate on the integral time scale with $M=1$, 
$T^I_{p,1}(n,t_u) \equiv \int_0^{t_u}F_p(n,t) dt$, 
the derivative time scale with $M=2$,
$T^D_{p,2} \equiv [\frac{\partial^2 F_p(n,t)}{\partial t^2}|_{t=0}]^{-1/2}$,
and the associated dynamic-multiscaling exponents defined via 
$T^I_{p,1}(n,t_u) \sim k_n^{-z^I_{p,1}}$ and 
$T^D_{p,2}(n) \sim k_n^{-z^D_{p,2}}$. In principle we should use
$t_u \rightarrow \infty$ but, since it is not possible to obtain $F_p(n,t)$
accurately for large $t$, we select an upper cut-off $t_u$ such that 
$F_p(n,t_u) = \alpha$, where, for all $n$ and $p$, we
choose $\alpha = 0.7$ in the results we report. We have
checked that our results do not change if we use $0.3 < \alpha < 0.8$. 
The slope of a log-log plot of $T^I_{p,1}(n)$ versus $k_n$ now
yields $z^I_{p,1}$ (Fig.~\ref{cofun} and Table~\ref{data-table}).  
Preliminary data for $z^I_{p,1}$ were reported by us in Ref.~\cite{mit03}.

For extracting the derivative scale $T^D_{p,2}$ we extend $F_p(n,t)$ to 
negative $t$ via $F_p(n,-t) = F_p(n,t)$ and use a centered, sixth-order, 
finite-difference scheme to find 
$\frac{\partial^2}{\partial t^2} F_p(n,t) |_{t=0}$. A log-log plot
of $T^D_{p,2}(n)$  versus $k_n$ now yields the exponent $z^D_{p,2}$
(Fig.~\ref{cofun} and Table~\ref{data-table}).

In Ref.~\cite{bif99} dynamic-multiscaling exponents were extracted not
from time-dependent structure functions but by using the following exit-time 
algorithm: 
We define the decorrelation time for shell $n$,
at time $t_i$, to be $T_i(n)$, such that,   
$|u_n(t_i)||u_n(t_i+T_i)| \ge \lambda^{\pm 1}|u_n(t_i)|^2$, 
with $ 0 < \lambda < 1$. The exit-time scale of order-$p$ and degree-$M$ 
for the shell $k_n$ is 
\begin{eqnarray} 
T^E_{p,M} \equiv \underset{N\rightarrow \infty}{\lim} 
              \biggl[\frac{\frac{1}{N}\sum_{i=1}^{N}T_i^M |u_n(t_i)|^p}
                       {\frac{1}{N}\sum_{i=1}^{N}|u_n(t_i)|^p}
               \biggl]
                  \sim k_n^{-z^{E}_{p,M}},
\label{texit}
\end{eqnarray}
 where the last proportionality follows from the dynamic-multiscaling ansatz.
In practice we cannot of course take the limit $N\rightarrow \infty$; in 
a typical run of length $T_{av}$ (Table~\ref{para-table}) $N \simeq 10^9$. 
By suitably adapting the multifractal formalism used above, we get the 
exit-time bridge relation
$z^E_{p,M} = 1 + [\zeta_{p-M} - \zeta_p]/M$,
obtained in Ref.~\cite{bif99} only for $M=1$. 
Dynamic-multiscaling exponents obtained via this exit-time algorithm 
are shown for $M=1$ and $M=-2$ in Table~\ref{data-table}.
The exit-time bridge relations for $M>0$ are the analogs of the 
integral-time bridge relation (\ref{zipm}) and those for $M<0$ are the 
analogs of the derivative-time bridge relation (\ref{zdpm}). 
We have checked that our results do not depend on $\lambda$ for 
$0.3<\lambda<0.8$.  

Our numerical results for the equal-time exponents $\zeta_p$ (Column 2),
the integral-time exponents $z^I_{p,1}$ (Columns 3 and 4), the 
derivative-time exponents $z^D_{p,2}$ (Columns 6 and 7), and the 
exit-time exponents $z^E_{p,1}$ and $z^E_{p,-2}$ (Columns 5
and 8, respectively) for $1\leq p \leq 6$ are given in Table~\ref{data-table}. 
The agreement of the exponents in Columns 3 and 4
shows that the bridge relation (\ref{zipm}) is satisfied (within
error bars). Likewise, a comparison of Columns 6 and 7 shows that 
the bridge relation (\ref{zdpm}) is satisfied. By comparing Columns 4 and 5 we 
see that the integral-time exponent $z^I_{p,1}$ is the same as the exit-time
exponent $z^E_{p,1}$; similarly, Columns 7 and 8 show that the
 derivative-time exponent $z^D_{p,2}$ is the same as the exit-time
exponent $z^E_{p,-2}$. The relation $z^D_{2,1} = \zeta_2$ mentioned 
above~\cite{hay00} is not meaningful in the GOY model since
$\partial F_p(n,t)/\partial t |_{t=0}$ vanishes, at least at the
 level of accuracy of our numerical study. 

We have obtained $50$ different values of each of the dynamic-multiscaling 
exponents from $50$ different initial conditions. For each of these initial
conditions time-averaging is done over a time $T_{av}$(Table~\ref{para-table}) 
which is larger than the averaging 
time of Ref.~\cite{bif99} by a factor of about $10^4$. The means of these
$50$ values for each of the dynamic-multiscaling exponents are shown in 
Table~\ref{data-table}; and the standard deviation 
yields the error. This averaging is another way of removing 
the effects of the 3-cycle mentioned above.  
\begin{table*}
\framebox{\begin{tabular}{c|c|c|c|c|c|c|c}
order$(p)$ & $\zeta_p$ & $z_{p,1}^I$[Eq.(\ref{zipm})] & $z_{p,1}^I$ &
$z^E_{p,1}$ & $z_{p,2}^D[Eq.(\ref{zdpm})]$ & $z_{p,2}^D$ & $z^E_{p,-2}$  \\
\hline
 1 &  0.3777 $\pm$ 0.0001 & 0.6221 $\pm$ 0.0001  & 0.60 $\pm$ 0.02   &
 0.603 $\pm$ 0.007 & 0.6820 $\pm$ 0.0001  & 0.70  $\pm$ 0.02  & 0.677 $\pm$0.001   \\
 2 &  0.7091 $\pm$ 0.0001 & 0.6686 $\pm$ 0.0002  & 0.67  $\pm$ 0.02  &  
 0.661$\pm$ 0.007 & 0.7081 $\pm$ 0.0002  & 0.71  $\pm$ 0.01 & 0.719 $\pm$0.004  \\
 3 &  1.0059 $\pm$ 0.0001 & 0.7030 $\pm$ 0.0002  & 0.701 $\pm$ 0.009& 
 0.708 $\pm$ 0.001& 0.7310 $\pm$ 0.0002  & 0.73  $\pm$ 0.01  & 0.739 $\pm$0.006  \\
 4 &  1.2762 $\pm$ 0.0002 &  0.7298 $\pm$ 0.0003 & 0.727 $\pm$ 0.007&
 0.74$\pm$0.01& 0.7509 $\pm$ 0.0003  & 0.744 $\pm$ 0.009 & 0.758 $\pm$0.006  \\
 5 &  1.5254 $\pm$ 0.0005 & 0.7511 $\pm$ 0.0007  & 0.759 $\pm$ 0.009  &  
 0.77$\pm$ 0.01 & 0.7684 $\pm$ 0.0007  & 0.756 $\pm$ 0.009   & 0.778 $\pm$0.003  \\
 6 &  1.757  $\pm$ 0.001  & 0.768  $\pm$ 0.002   & 0.77  $\pm$ 0.01  & 
 0.79$\pm$ 0.01 & 0.7836 $\pm$ 0.002   & 0.764 $\pm$ 0.009  & 0.797 $\pm$0.0008   \\
\end{tabular}}
\caption{Order$-p\/$ (Column 1) multiscaling exponents for $1\leq p \leq6$ 
from our simulations of the GOY model: equal-time exponents $\zeta_p$
(Column 2), integral-scale dynamic-multiscaling exponent 
$z^I_{p,1}\/$ of degree-$1$ (Column 3) from the bridge relation (\ref{zipm}) 
and the values of $\zeta_p$ in Column 1, $z^I_{p,1}$ from our 
calculation using time-dependent structure functions (Column 4), the exit-time 
exponents of order $1$ $z^E_{p,1}$ (Column 5), the derivative-time exponents 
$z^D_{p,2}$ (Column 6) from the bridge relation (\ref{zdpm}) and the values of 
$\zeta_p$ in Column 1, $z^D_{p,2}$ from our calculation 
using time-dependent structure function (Column 7) and the exit-time
exponent of order $-2$, $z^E_{p,-2}$ (Column 8). The error estimates are 
obtained as described in the text.}
\label{data-table}
\end{table*}

We have shown systematically how different ways of extracting 
time scales from time-dependent velocity structure functions or time series 
can lead to different sets of dynamic-multiscaling exponents, which are 
related in turn to the equal-time multiscaling exponents $\zeta_p$ by 
different classes of bridge relations. Our extensive numerical study of the 
GOY shell model for fluid turbulence verifies explicitly that such bridge 
relations hold. Experimental studies of Lagrangian quantities 
in turbulence have been increasing over the past few years~\cite{lag}. We hope
our work will stimulate studies of dynamic multiscaling in such
experiments. Furthermore, the sorts of bridge relations we have discussed
here must also hold in other problems with multiscaling of equal-time
and time-dependent structure functions or correlation functions. 
Passive-scalar and magnetohydrodynamic turbulence are 
two obvious examples which we will report on elsewhere \cite{mit}. Numerical 
studies of time-dependent, quasi-Lagrangian-velocity structure functions in 
the Navier-Stokes equation, already under way, will also be discussed 
elsewhere.

We thank A. Celani, S.K. Dhar, U. Frisch, S. Ramaswamy,  
 A. Sain, and especially C. Jayaprakash for discussions. 
This work was supported by the Indo-French Centre for the Promotion 
of Advanced Research (IFCPAR Project No. 2404-2).  
D.M. thanks the Council of Scientific and Industrial Research, 
India for support.

\printfigures
\printtables

\end{document}